\documentclass[aps, prb, twocolumn, superscriptaddress, floatfix, 10pt]{revtex4-2}
\usepackage{amsmath, amsfonts, amssymb, graphicx, bm}
\usepackage{multirow, color, textcomp, subfigure, nicefrac, float, enumitem, cancel, xcolor, stmaryrd, wasysym}
\usepackage{url}
\usepackage[colorlinks=true,citecolor=blue,urlcolor=blue,linkcolor=blue]{hyperref}
\usepackage{arydshln}
\usepackage{kotex}
\usepackage[normalem]{ulem}

\newcommand{\R}[1]{\textcolor{red}{#1}}

\hypersetup{colorlinks=true, urlcolor=blue, citecolor=blue, pdfborder={0 0 0},}

\begin{document}

\title{Unconventional p-wave and finite-momentum superconductivity induced by altermagnetism through the formation of Bogoliubov Fermi surface}

\author{SeungBeom Hong}
\affiliation{Department of Physics, Hanyang University, Seoul 04763, Republic of Korea}

\author{Moon Jip Park}
\email{moonjippark@hanyang.ac.kr}
\affiliation{Department of Physics, Hanyang University, Seoul 04763, Republic of Korea}
\affiliation{Research Institute for Natural Science, Hanyang University, Seoul 04763, Korea}

\author{Kyoung-Min Kim}
\email{kmkim@ibs.re.kr}
\affiliation{Center for Theoretical Physics of Complex Systems, Institute for Basic Science, Daejeon 34126, Republic of Korea}

\date{\today}

\begin{abstract}
Altermagnet is an exotic class of magnetic materials wherein the Fermi surface exhibits a momentum-dependent spin-splitting while maintaining a net zero magnetization. Previous studies have shown that this distinctive spin-splitting can induce chiral $p$-wave superconductors or Fulde-Ferrell (FF) superconducting states carrying finite momentum. However, the underlying mechanisms of such unconventional superconductivities remain elusive. Here, we propose that the formation of the Bogoliubov Fermi surface (BFS) through the exchange field can play a significant role in such phenomena. Through a systematic self-consistent mean-field analysis on the extended attractive Hubbard model combined with the $d$-wave spin-splitting induced by the exchange field, as observed in RuO\textsubscript{2}, we demonstrate that the formation of the BFS suppresses conventional spin-singlet superconducting states with $s$-wave characteristics. In contrast, the chiral $p$-wave state maintains a fully gapped spectrum without the Fermi surface, thereby becoming the ground state in the strong field regime. In the intermediate regime, we find that the FF state becomes the predominant state through the optimization of available channels for Cooper pairing. Moreover, we illustrate how the prevalence of the chiral $p$-wave and FF states over the $s$-wave state changes under the variation of the field strength or chemical potential. Our findings provide valuable insights into potential pathways for realizing sought-after topological $p$-wave superconductivity and finite momentum pairing facilitated by altermagnetism.
\end{abstract}

\maketitle

\begin{figure*}[t!]
    \centering
    \includegraphics[width=0.8\linewidth]{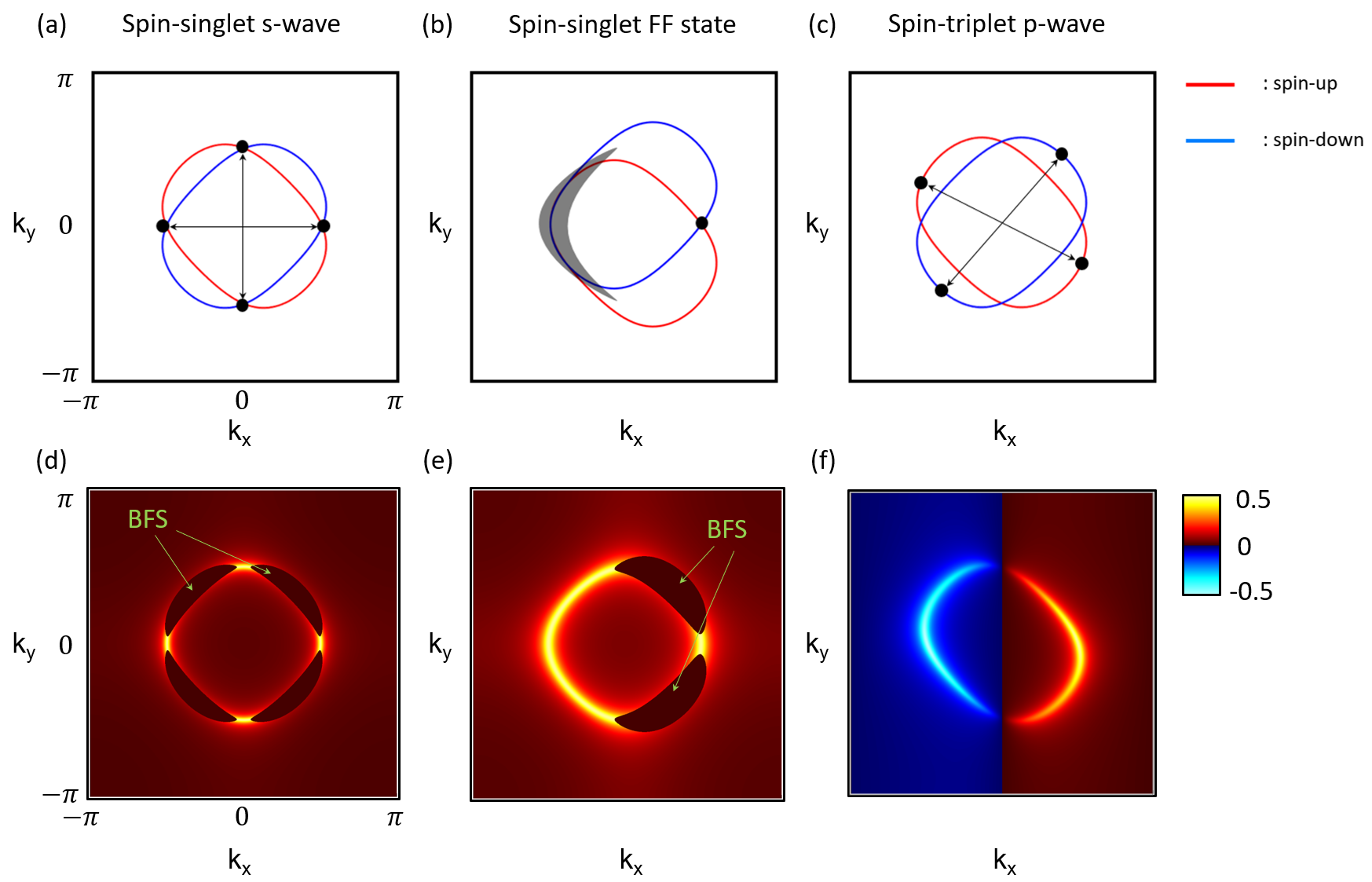}
    \caption{(a--c) Three predominant superconducting states in altermagnetic metals: (a) the spin-singlet $s$-wave state (b) spin-singlet Fulde-Ferrell (FF) state with a finite-momentum $s$-wave pairing, and (c) the spin-triplet $p$-wave state. (a) In the $s$-wave state, Cooper pairs are formed exclusively from electrons at nodal points, represented by black dots, where the spin-up Fermi surface (FS) overlaps with the spin-down FS, each represented by red and blue, respectively. (b) In the FF state, Cooper pairs form not only at the isolated nodal point but also across a broader region, as indicated by the gray-shaded area, where the two FSs overlap. Here, the spin-up FS and spin-down FS are characterized by $(\bm{k}+\bm{q})$ and $(-\bm{k}+\bm{q})$, respectively, where $\bm{q}$ represents the pairing momentum. (c) In the $p$-wave state, two electrons from the same FS form Cooper pairs, as indicated by two black dots in each FS. Consequently, the formation of Cooper pairs occurs across the entire FS, not isolated points, in contrast to (a). (d–f) The pair correlation functions for (d) the $s$-wave state, (e) the FF state, and (f) the $p$-wave state are shown. In panels (d) and (e), the shaded areas indicate where the correlation function vanishes due to the formation of the Bogoliubov Fermi surface (BFS) in these local regions. In contrast, panel (f) shows an absence of the BFS, with the correlation function remaining consistent across the entire Fermi surface.} 
    \label{fig1}
\end{figure*}

\section{Introduction}

Breaking time-reversal symmetry (TRS) offers pathways to explore unconventional superconductivity beyond the conventional Bardeen–Cooper–Schrieffer (BCS) paradigm by inducing spin-split Fermi surfaces. While electrons encounter obstacles in forming spin-singlet Cooper pairs within the conventional framework \cite{RevModPhys.63.239}, the spin-split Fermi surfaces may accommodate unconventional Cooper pairs with distinctive pairing symmetries. Noteworthy examples include $p$-wave topological superconductivity, which can be induced by external magnetic fields or internal exchange fields within ferromagnetic metals \cite{RevModPhys.83.1057, Sato_2017, doi:10.7566/JPSJ.85.072001} or even by inversion symmetry breaking perturbation that breaks `effective time-reversal symmetry' in antiferromagnetic metals \cite{PhysRevLett.126.067001}. Notably, these scenarios often imply the existence of intriguing Majorana edge or corner modes. Alternatively, electrons can form finite-momentum pairings exhibiting nonuniform modulation in their pairing amplitudes, leading to the emergence of Fulde–Ferrell–Larkin–Ovchinnikov (FFLO) states \cite{https://doi.org/10.1002/andp.201700282,doi:10.1143/JPSJ.76.051005,PhysRevLett.108.117003, doi:10.1073/pnas.2019063118, Chen2018}. Moreover, TRS breaking can lead to a Fermi surface with gapless Bogoliubov quasiparticles, known as a Bogoliubov Fermi surface (BFS), where superconductivity can still persist due to the coherence of electron pairs, even without a fully gapped spectrum \cite{PhysRevB.101.024505, PhysRevLett.118.127001, PhysRevB.102.064504, PhysRevB.57.8566}.

Recently, considerable attention has been drawn to the impact of distinct TRS breaking patterns on superconductivity, stemming from a novel collinear magnetism known as altermagnetism \cite{PhysRevLett.125.147001, PhysRevLett.132.166702, Kebler2024, PhysRevB.108.224421, PhysRevB.110.205120, PhysRevB.109.134511, PhysRevLett.131.076003, hu2024nonlinearsuperconductingmagnetoelectriceffect, PhysRevB.109.024517, PhysRevB.109.134511, PhysRevB.108.205410, PhysRevB.108.184505, PhysRevLett.133.106601, PhysRevB.108.L060508, PhysRevB.108.054511, PhysRevB.110.L060508, Zhang2024, PhysRevB.110.L140506, PhysRevB.109.L220505, PhysRevB.109.L201404, PhysRevB.110.094508, chourasia2024thermodynamicpropertiessuperconductorinterfaced, PhysRevB.109.134515, PhysRevB.110.024503, PhysRevB.110.014518, chakraborty2024perfectsuperconductingdiodeeffect}. In such systems, due to symmetry considerations, the Fermi surface displays momentum-dependent spin-splitting while maintaining a net magnetization of zero overall \cite{PhysRevB.107.L100418, PhysRevLett.132.036702, Reimers2024, PhysRevB.99.184432, doi:10.1126/sciadv.aaz8809}. A prominent illustration of such an altermagnetic spin-splitting effect is the case of a two-dimensional altermagnetic metal like RuO\textsubscript{2}, wherein the Fermi surface manifests $d$-wave-like spin splitting with a four-fold symmetric point nodal degeneracy. Remarkably, under applied strain, thin film RuO$_2$ exhibits superconductivity at relatively high temperatures, with $T_c=1.7$ K \cite{PhysRevLett.125.147001}, suggesting it as a promising platform for investigating the influence of such distinct spin-split Fermi surfaces on superconductivity. However, recent $\mu$SR experiments have reported the absence of magnetic moments in RuO\textsubscript{2}, challenging its classification as an altermagnet \cite{PhysRevLett.132.166702, Kebler2024}. Previous theoretical studies have suggested intriguing phenomena related to superconductivity, including unconventional superconductivity \cite{PhysRevB.110.205120, PhysRevB.108.224421}, Josephson junction effect \cite{PhysRevLett.131.076003, hu2024nonlinearsuperconductingmagnetoelectriceffect, PhysRevB.109.024517, PhysRevB.109.134511}, topological superconductivity with Majorana edge modes \cite{PhysRevB.108.205410, PhysRevB.108.184505, PhysRevLett.133.106601}, orientation-dependent Andreev reflection \cite{PhysRevB.108.L060508, PhysRevB.108.054511}, finite-momentum Cooper pairing \cite{PhysRevB.110.L060508,  Zhang2024, PhysRevB.110.L140506, PhysRevB.109.L220505}, gapless superconductivity \cite{PhysRevB.109.L201404}, thermoelectric effect \cite{PhysRevB.110.094508, chourasia2024thermodynamicpropertiessuperconductorinterfaced}, many body magnonic effect \cite{PhysRevB.109.134515}, and superconducting diode effect \cite{PhysRevB.110.024503, PhysRevB.110.014518, chakraborty2024perfectsuperconductingdiodeeffect}. Despite significant recent advances, the impact of the altermagnetic spin-splitting effect on observed superconductivity in these systems remains incompletely understood.

In this Letter, we explore the emergence of unconventional superconducting states resulting from altermagnetic spin-splitting of the Fermi surface. Through a systematic mean-field analysis on the extended attractive Hubbard model \cite{PhysRevB.108.064514} incorporating $d$-wave altermagnetic exchange fields, we uncover a spectrum of unconventional superconducting phenomena alongside conventional $s$-wave BCS states. Over a wide parameter range, we observe that the exchange fields suppress the $s$-wave BCS state by reducing the corresponding $s$-wave channel correlation function for Cooper pair formation, attributed to the formation of BFS (Fig.~\ref{fig1}(a)). Conversely, topological $p+ip$ superconductivity is favored as the pertinent $p$-wave channel correlation function remains robust against the exchange field (Fig.~\ref{fig1}(c)). Furthermore, we demonstrate that $s$-wave superconducting states with finite momentum pairing and accompanying BFS, namely `gapless FFLO states,' may emerge as intermediate states preceding $p+ip$ states due to the augmentation of the available channels for Cooper pairing (Fig.~\ref{fig1}(b)). While topological chiral $p$-wave states \cite{PhysRevB.108.184505} and finite-momentum pairing states \cite{PhysRevB.110.L060508, Zhang2024} have been reported, our investigation demonstrates that their emergence can be intrinsically linked to the formation of BFS. Furthermore, we investigate the topological gap structures for different pairing symmetry. We discuss the possible experimental detection to distinguish the pairing symmetry using London penetration depth measurement. These diverse manifestations highlight the symmetry-enriched unconventional superconductivity phenomena \cite{Pal_2024, PhysRevB.110.104515} arising from the time-reversal symmetry breaking of altermagnetism. \\

\section{Model}

\subsection{Extended attractive Hubbard model incorporating $d$-wave altermagnetic exchange fields}

We consider two-dimensional metallic systems possessing $d$-wave altermagnetism, as observed in RuO\textsubscript{2} \cite{PhysRevB.99.184432, doi:10.1126/sciadv.aaz8809}. The model can be described using an effective tight-binding Hamiltonian given by

\begin{equation}
\begin{aligned}
    \hat{H} = & \; -t\sum_{\langle i,j\rangle, \sigma} c_{i\sigma}^\dagger c_{j\sigma} - \mu\sum_{i, \sigma}  c_{i\sigma}^\dagger c_{i\sigma} \\
    & - \frac{1}{4}\sum_{\langle\langle i,j\rangle\rangle,\sigma} s(\sigma) J_{i,j} c_{i,\sigma}^\dagger c_{j,\sigma} \\
    & - U\sum_{i}n_{i\uparrow}n_{i\downarrow} - V\sum_{\sigma,\sigma'}\sum_{\langle i,j \rangle}n_{i\sigma}n_{j\sigma'}.
     \label{eq:H_full_main}
\end{aligned}
\end{equation}
Here, $c^\dagger_{i\sigma}$ and $n_{i\sigma}=c^\dagger_{i\sigma}c_{i\sigma}$ represent the electron creation and density operators at the $i$-site with the total lattice number $N$, spin indices $\sigma=\uparrow,\downarrow$, projected in the altermagnetic band. $c^{\dagger}_{\bm{k}\sigma}$ is the electron creation operator in the momentum space. The parameters $t$ and $\mu$ represent the hopping between nearest-neighbor sites and the chemical potential, respectively. The exchange field $J_{i,j}$ induces a $d$-wave-like spin splitting, characterized by the following sign alternation pattern: $J_{i,j}=J$ for $\bm{r}_{ij}=\hat{x}+\hat{y}$ and $\bm{r}_{ij}=-\hat{x}-\hat{y}$, while $J_{i,j}=-J$ for $\bm{r}_{ij}=\hat{x}-\hat{y}$ and $\bm{r}_{ij}=-\hat{x}+\hat{y}$. Here, $\bm{r}_{ij}=\bm{r}_i-\bm{r}_j$ represents the coordinate displacement between the next-nearest neighbor $i$ and $j$-sites, and $J$ signifies the strength of the exchange field. $U>0$ and $V>0$ represent the phonon-induced attractive onsite and nearest-neighbor density-density interactions, respectively. We denote $s(\uparrow)=+1$ and $s(\downarrow)=-1$. The Bloch Hamiltonian, $h_{\bm{k}}=\xi_{\bm{k}}\textrm{I}_2+J_{\bm{k}}\sigma_z $ with $\xi_{\bm{k}} = -2t \left[ \cos (k_x)+\cos (k_y)\right]-\mu$ and $J_{\bm{k}} = J\sin (k_x)\sin (k_y)$, respects $C_{4z}\mathcal{T}$ by satisfying $\sigma_{y}h^{*}(k_x,k_y)\sigma_y=h(k_y,-k_x)$. The Fermi surface has the four nodal point degeneracies protected by $C_{4z}\mathcal{T}$ despite the spin splitting in the Fermi surface (Fig.~\ref{fig1}(a)). We explore a wide range of values for $J$, $\mu$, and $V$: $0\le J/2t \le 1$, $-2\le \mu/2t\le 0$, and $0\le V/2t \le 1.5$, while maintaining a fixed value of $U/2t=1.5$. It is important to note that the real-space model requires more than two sublattice degrees of freedom \cite{PhysRevB.110.144412, PhysRevB.108.224421}. The interaction terms written in the lattice model would be modified upon the projection to the low-energy effective model. Nevertheless, in this work, we investigate the role of the on-site and nearest-neighbor interactions based on the single-band Fermi surface.

To investigate the emergence of unconventional superconducting states, we conduct a self-consistent mean-field analysis utilizing the Bogoliubov-de-Gennes (BdG) formalism. Due to the presence of the inversion symmetry, the spin-singlet and triplet states remain uncoupled, allowing separate consideration. The following five distinct superconducting states may appear in our model \cite{PhysRevB.108.184505, PhysRevB.108.064514}: (i) $s$-wave spin-singlet (ii) Fulde-Ferrell (FF) spin-singlet state, originating from $U$, and the remaining three from $V$, which include (iii) extended $s$-wave spin-singlet, (iv) $d$-wave spin-singlet, (v) $p$-wave spin-triplet states. In the following, we focus on three predominant states: the $s$-wave state, the FF state, and the $p$-wave state. These states are prevalent within the parameter space under consideration. Detailed analysis for the extended $s$-wave and $d$-wave states can be found from the Supplementary Material (SM) \cite{SM}.

\subsection{Mean-field analysis using BdG formalism}

We consider the following three pairing amplitudes:
\begin{equation}
\begin{aligned}
    \Delta^{\textrm{s}}_{\uparrow\downarrow} & = - \frac{U}{N} \sum_{\bm{k}}  \left \langle c_{\bm{k}\uparrow}c_{-\bm{k}\downarrow} \right\rangle, \\
    \Delta^{\textrm{FF}}_{\uparrow\downarrow; \bm{q}} & = - \frac{U}{N} \sum_{\bm{k}} \left \langle c_{\bm{k+q}\uparrow}c_{\bm{-k+q}\downarrow} \right\rangle, \\
    \Delta^{\textrm{p+ip}}_{\sigma\sigma} & = - \frac{V}{N} \sum_{\bm{k}} (g^\textrm{p+ip}_{\bm{k}})^{*}\left \langle c_{\bm{k}\sigma}c_{-\bm{k}\sigma} \right\rangle, \label{eq:Delta_def}
\end{aligned}
\end{equation}
to characterize the $s$-wave ($\Delta^{\textrm{s}}_{\uparrow\downarrow}$), FF ($\Delta^{\textrm{FF}}_{\uparrow\downarrow;\bm{q}}$), and $p$-wave ($\Delta_{\sigma\sigma}^{\textrm{p+ip}}$) states. The FF state forms the spin singlet, and it features spatial modulation characterized by a finite wave vector $\bm{q}$. Here, $g^\textrm{p+ip}_{\bm{k}} = \sin(k_x) + i \sin(k_y)$ represents a square harmonics exhibiting $p_x+ip_y$-like characteristics. As such, we refer to this $p$-wave state as a $p+ip$ state \cite{PhysRevB.108.184505}. The BdG quasiparticle energy spectra of these states exhibit distinct gap-opening patterns, as shown by
\begin{equation}
\begin{aligned}
    E^{\textrm{s}}_{\bm{k}\sigma} & = \sqrt{\xi_{\bm{k}}^{2}+|\Delta^\textrm{s}_{\uparrow\downarrow}|^{2}} + s(\sigma) J_{\bm{k}}, \\
    E^{\textrm{FF}}_{\bm{k}\sigma;\bm{q}} & = \sqrt{\Xi_{\bm{k};\bm{q}}^{2} + |\Delta^{\textrm{FF}}_{\uparrow\downarrow; \bm{q}}|^{2}} + s(\sigma) \Lambda_{\bm{k};\bm{q}}, \\\
    E^{\textrm{p+ip}}_{\bm{k}\sigma} & = \sqrt{(\xi_{\bm{k}} + s(\sigma) J_{\bm{k}})^{2}+|\Delta_{\sigma\sigma}^{\textrm{p+ip}}g_{\bm{k}}^{\textrm{p+ip}}|^{2}},  \label{eq:BdG_en}
\end{aligned}
\end{equation}
where $E^{\textrm{s}}_{\bm{k}\sigma}$ corresponds to the $s$-wave state, $E^{\textrm{FF}}_{\bm{k}\sigma;\bm{q}}$ to the FF state, and $E^{\textrm{p+ip}}_{\bm{k}\sigma}$ to the $p+ip$ state. The functions $\Xi_{\bm{k};\bm{q}}=\frac{1}{2}(\xi_{\bm{k}+\bm{q}}+J_{\bm{k}+\bm{q}}+\xi_{\bm{k}-\bm{q}}-J_{\bm{k}-\bm{q}})$ and $\Lambda_{\bm{k};\bm{q}}=\frac{1}{2}(\xi_{\bm{k}+\bm{q}}+J_{\bm{k}+\bm{q}}-\xi_{\bm{k}-\bm{q}}+J_{\bm{k}-\bm{q}})$ describe the modified dispersions for the FF phase with finite momentum $\bm{q}$. The energies $E^{\textrm{s}}_{\bm{k}\sigma}$ and $E^{\textrm{FF}}_{\bm{k}\sigma;\bm{q}}$ can be negative value under strong exchange field strength $J_{\bm{k}}$. In contrast, the gap $E^{\textrm{FF}}_{\bm{k}\sigma;\bm{q}}$ of the $p+ip$ state maintains its gapped nature regardless of the field strength.

The pairing amplitudes, as presented in Eq.~\eqref{eq:Delta_def}, are self-consistently determined solving the following gap equations at zero temperature ($T=0$):
\begin{equation}
\begin{aligned}
    \Delta^\textrm{s}_{\uparrow\downarrow} & = \frac{U}{N} \sum_{\bm{k}} \frac{\Delta^\textrm{s}_{\uparrow\downarrow}}{2\epsilon_{\bm{k}}^{\textrm{s}}}\big[1-\Theta(-E_{\bm{k}\uparrow}^\textrm{s})-\Theta(-E_{\bm{k}\downarrow}^{s})\big],  \\
    \Delta^{\textrm{FF}}_{\uparrow\downarrow; \bm{q}} & = \frac{U}{N} \sum_{\bm{k}} \frac{\Delta^{\textrm{FF}}_{\uparrow\downarrow; \bm{q}}}{2\epsilon^{\textrm{FF}}_{\bm{k};\bm{q}}} \big[1 - \Theta(-E^{\textrm{FF}}_{\bm{k}\uparrow;\bm{q}}) - \Theta(-E^{\textrm{FF}}_{\bm{k}\downarrow;\bm{q}})\big], \\
    \Delta_{\sigma\sigma}^\textrm{p+ip} & = \frac{V}{N} \sum_{\bm{k}} \frac{\Delta_{\sigma\sigma}^{\textrm{p+ip}}|g^\textrm{p+ip}_{\bm{k}}|^{2}}{2E^{\textrm{p+ip}}_{\bm{k}\sigma}}. \label{eq:gap_eq}
\end{aligned}
\end{equation}
In these equations, we denote $\epsilon_{\bm{k}}^{\textrm{s}}=\sqrt{\xi_{\bm{k}}^{2}+|\Delta^{\textrm{s}}_{\uparrow\downarrow}|^2}$ and $\epsilon_{\bm{k};\bm{q}}^{\textrm{FF}}=\sqrt{\Xi_{\bm{k};\bm{q}}^{2} + |\Delta^{\textrm{FF}}_{\uparrow\downarrow; \bm{q}}|^{2}}$. $\Theta(E)$ is the Heaviside step function. The gap equations for the $s$-wave and FF states incorporate form factors $\big[1-\Theta(-E^{\textrm{s}}_{\bm{k}\uparrow})-\Theta(-E^{\textrm{s}}_{\bm{k}\downarrow})\big]$ and $\big[1 - \Theta(-E^{\textrm{FF}}_{\bm{k}\uparrow;\bm{q}}) - \Theta(-E^{\textrm{FF}}_{\bm{k}\downarrow;\bm{q}})\big]$, respectively. These factors exclude contributions from occupied BdG quasiparticle states with negative energies ($E^{\textrm{s}}_{\sigma\bm{k}}<0$ or $E^{\textrm{FF}}_{\bm{k}\sigma;\bm{q}}<0$). Moreover, the form factors reflect the formation of BFS in the singlet quasiparticle spectrum (zero excitation energy) and zero pair correlation region (negative excitation energy) in Fig. \eqref{fig1} (d, e). Detailed analysis is provided in SM \cite{SM}. Here, the physical definition of the $s$-wave and FF state BFS, $\bm{k}^{\textrm{s},\sigma}_{\textrm{BFS}}$, $\bm{k}^{\textrm{FF},\sigma}_{\textrm{BFS}}$ is the set of momentum $\bm{k}$ that has zero excitation energy:
\begin{equation}
    \begin{aligned}
        \{\bm{k}^{\textrm{s},\sigma}_{\textrm{BFS}}\}&=\{\bm{k}|E^{\textrm{s}}_{\bm{k}\sigma}=0\},\\
        \{\bm{k}^{\textrm{FF},\sigma}_{\textrm{BFS}}\}&=\{\bm{k}|E^{\textrm{FF}}_{\bm{k}\sigma}=0\}.
    \end{aligned}
\end{equation}
Conversely, $\Delta^{\textrm{p+ip}}_{\sigma\sigma}$ for the $p+ip$ state incorporates contributions from all BdG quasiparticle states. For the temperature dependence of the Eq. \eqref{eq:gap_eq}, detailed forms are shown in Eq. (35), (64), (81) of the SM \cite{SM}.

The ground state energy for each state is calculated using the following energy expressions:
\begin{equation}
\begin{aligned}
    F^{\textrm{s}}_{0} = & \; \sum_{\bm{k}}\Bigg[\xi_{\bm{k}} - \epsilon_{\bm{k}}^{\textrm{s}} + \frac{|\Delta^{\textrm{s}}_{\uparrow\downarrow}|^2}{U}\Bigg] +\sum_{\bm{k},\sigma}\Theta(-E^{\textrm{s}}_{\bm{k}\sigma})E^{\textrm{s}}_{\bm{k}\sigma}, \\
    F^{\textrm{FF}}_{0}(\bm{q}) = & \; \sum_{\bm{k}}\Bigg[\Xi_{\bm{k};\bm{q}} - \epsilon_{\bm{k};\bm{q}}^{\textrm{FF}} + \frac{|\Delta^{\textrm{FF}}_{\uparrow\downarrow;\bm{q}}|^2}{U}\Bigg]  \\
    & + \sum_{\bm{k},\sigma}\Theta(-E^{\textrm{FF}}_{\bm{k}\sigma;\bm{q}})E^{\textrm{FF}}_{\bm{k}\sigma;\bm{q}}, \\
    F^{\textrm{p+ip}}_{0} = & \; \frac{1}{2} \sum_{\bm{k},\sigma} \Bigg[ \xi_{\bm{k}}-E^{\textrm{p+ip}}_{\bm{k}\sigma} + \frac{|\Delta_{\sigma\sigma}^{\textrm{p+ip}}|^2}{V} \Bigg], \label{eq:gr_en}
\end{aligned}
\end{equation}
where $F^{\textrm{s}}_{0}$, $F_{0}^{\textrm{FF}}(\bm{q})$, and $F_{0}^\textrm{p}$ correspond to the $s$-wave, FF, and $p+ip$ states, respectively. For the $s$-wave and FF states, the ground state energy incorporates contributions from the occupied BdG quasiparticle states with negative energy ($E^{\textrm{s}}_{\bm{k}\sigma}<0$ or $E^{\textrm{FF}}_{\bm{k}\sigma;\bm{q}}<0$) in addition to the usual energy contribution in the square brackets. In contrast, the ground state energy for the $p+ip$ state only contains the usual energy contribution due to its positive definite nature ($E^{\textrm{p+ip}}_{\bm{k}\sigma}>0$). In the FF state, we identify the optimal $\bm{q}$ vector by evaluating $F^{\textrm{FF}}_{0}(\bm{q})$ across various $\bm{q}$ values and selecting the specific $\bm{q}$-value that yields the minimum of $F^{\textrm{FF}}_{0}(\bm{q})$ (refer to Fig.~\ref{fig3}(a–b)). Refer to the SM \cite{SM} for the derivation of the gap equations and ground state energy expressions. \\

\begin{figure*}[t!]
    \centering
    \includegraphics[width=1\linewidth]{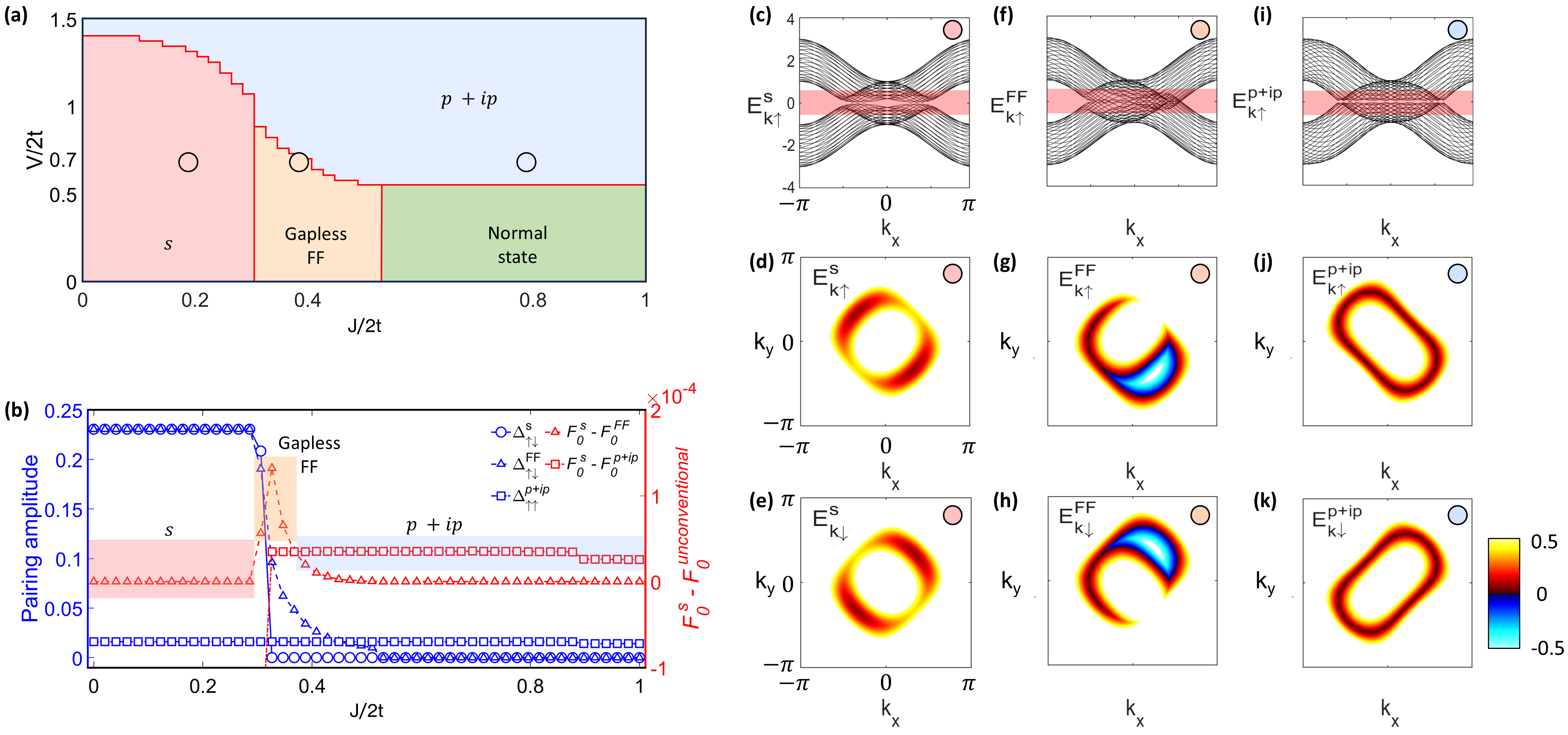}
    \caption{(a) Zero temperature phase diagram as a function of the strength of the exchange field ($J$) and the attractive nearest-neighbor density-density interaction ($V$), divided by the nearest-neighbor hopping strength ($t$). Three superconducting phases are found: the $s$-wave phase (denoted by `$s$'), the FF phase (`gapless FF'), and the $p+ip$ phase (`$p+ip$'), in addition to a non-superconducting metallic state (`Normal state'). (b) The evolution of the three superconducting states as a function of $J/2t$ for a fixed value of  $V/2t=0.7$. The left axis illustrates the pairing amplitudes for each state ($\Delta_{\uparrow\downarrow}^\textrm{s}$ for the $s$-wave, $\Delta_{\uparrow\downarrow}^\textrm{FF}$ for the FF phase, and $\Delta_{\uparrow\uparrow}^{\textrm{p}+\textrm{ip}}$ for the $p+ip$ phase). The right axis shows the differences in the ground state energy per site, specifically ($F_0^\textrm{s}-F_0^\textrm{FF}$) between the $s$-wave and FF states, and ($F_0^\textrm{s}-F_0^{\textrm{p}+\textrm{ip}}$) between the $s$-wave and $p+ip$ states. The energetically favored state within each $J/2t$ range is highlighted with distinct colors. (c--k) Bogoliubov-de-Gennes (BdG) quasiparticle energy spectra for (c--e) the $s$-wave state ($E_{\bm{k}\sigma}^\textrm{s}$), (f--h) the FF state ($E_{\bm{k}\sigma}^\textrm{FF}$), and (i--k) the $p+ip$ state ($E_{\bm{k}\sigma}^{\textrm{p}+\textrm{ip}}$). Panels (c, f, i) present the full energy spectra for the up spin sector ($\sigma=\uparrow$) as a function of $k_x$, with different curves corresponding to various $k_y$ values. Panels (d, g, j) display the BdG quasiparticle energy spectra for the up spin sector ($\sigma = \uparrow$) as a function of $(k_x, k_y)$, within the energy window highlighted in red in panels (c, f, i). Panels (e, h, k) showcase the BdG quasiparticle energy spectra for the down spin sector ($\sigma=\downarrow$) within the same energy window. In all panels (d--e, g--h, j--k), the color scale represents the values of the energy levels. Additionally, in panels (g--h), negative energy levels shown in blue indicate the formation of the BFS. The parameters for the on-site density-density interaction ($U$) and the chemical potential ($\mu$) are set as follows: $U/2t=1.5$ and $\mu/2t=-1$. The values for $J/2t=0.2$, $J/2t=0.4$, and $J/2t=0.8$ are employed in (c--e), (f--h), and (i--k), respectively, while $V/2t=0.7$ is consistently used throughout, as indicated by the circles in panel (a).} \label{fig2}
\end{figure*}

\section{Results}

\subsection{Zero temperature superconducting phase diagram}

We determine the energetically favored superconducting states across a range of parameters $J/2t$, $V/2t$, and $\mu/2t$, by numerically solving the gap equations (Eq.~\eqref{eq:gap_eq}) and computing the ground state energy for each state (Eq.~\eqref{eq:gr_en}). Fig.~\ref{fig2}(a) plots the resulting phase diagram as a function of $J/2t$ and $V/2t$, showing three superconducting phases, the $s$-wave, FF, and $p+ip$ phases, each having the lowest energy in the respective region. We observe that the $s$-wave phase (the red region in Fig.~\ref{fig2}(a)) predominates in the small exchange field regime ($J/2t \lesssim 0.28$). In this phase, the BdG quasiparticle energy spectra, $E^{\textrm{s}}_{\bm{k}\sigma}=\sqrt{\xi_{\bm{k}}^{2}+|\Delta^\textrm{s}_{\uparrow\downarrow}|^{2}} + s(\sigma) J_{\bm{k}}$, exhibit a fully gapped structure in both spin sectors $\sigma=\;\uparrow, \downarrow$ (Fig.~\ref{fig2}(c)). However, the exchange field $J_{\bm{k}}$ lifts their degeneracy, resulting in anisotropic, sign-alternating energy shifts (Fig.~\ref{fig2}(d–e)). In $E^{\textrm{s}}_{\bm{k}\uparrow}$, the shifted energy $J_{\bm{k}}$ is negative along the line $k_x=-k_y$ or positive along $k_x=k_y$ (Fig.~\ref{fig2}(d)). The magnitude of $J_{\bm{k}}$ reaches at its maximum value $J_* = J\sin^{2}(k_*)$, at $(k_x, k_y)=\pm(k_*,-k_*)$ where $k_*=\arccos{\Big(\frac{-(\mu-J)}{2t+\sqrt{(2t)^2-J(\mu-J)}}\Big)}$. At this point, $J_{\bm{k}}$ effectively reduces the energy gap as $\Delta^{\textrm{s}}_{\uparrow\downarrow}-J_*$, which aligns with the actual gap size $\bar{\Delta}_{\textrm{band}}$ in the energy spectra (e.g., $\Delta^{\textrm{s}}_{\uparrow\downarrow}=0.2296$, $J_*=0.1355$, and $\bar{\Delta}_{\textrm{band}}=0.0789 \approx \Delta^{\textrm{s}}_{\uparrow\downarrow}-J_*$ for the energy spectra presented in Fig.~\ref{fig2}(c)). It is important to note that as long as $J_*$ does not exceed $\Delta^{\textrm{s}}_{\uparrow\downarrow}$, $E^{\textrm{s}}_{\bm{k}\sigma}$ can maintain their full gap structures, ensuring that the $s$-wave state remains the ground state of the system. The opposite pattern is observed in $E^{\textrm{s}}_{\bm{k}\downarrow}$ (Fig.~\ref{fig2}(e)).

As $J$ further increases to $J/2t = 0.28$, $J_*$ starts to exceed $\Delta^{\textrm{s}}_{\uparrow\downarrow}$, leading to the gap closing and the formation of BFS in $E^{\textrm{s}}_{\bm{k}\sigma}<0$ in the $s$-wave state. We note that the formation of the BFS indicates the exclusion of contributions from these occupied BdG quasiparticle states in the self-consistency relation of $\Delta^{\textrm{s}}_{\uparrow\downarrow}$ (refer to the corresponding gap equation in Eq.~\eqref{eq:gap_eq}). Consequently, $\Delta^{\textrm{s}}_{\uparrow\downarrow}$ is significantly suppressed, leading to a considerable increase in the energy of the $s$-wave state \R{\cite{PhysRevB.57.8566, PhysRevB.102.064504}} (Fig.~\ref{fig2}(b)). Hence, the $s$-wave state is disfavored in this regime. Alternatively, the FF state becomes the most energetically favored state in this regime (the orange region in Fig.~\ref{fig2}(a)), which possesses a more pronounced pairing amplitude and reduced energy compared to the $s$-wave state (Fig.~\ref{fig2}(b)). The BdG quasiparticle energy spectra of the FF state, $E^{\textrm{FF}}_{\bm{k}\sigma;\bm{q}} = \sqrt{\Xi_{\bm{k};\bm{q}}^{2} + |\Delta^{\textrm{FF}}_{\uparrow\downarrow; \bm{q}}|^{2}} + s(\sigma) \Lambda_{\bm{k};\bm{q}}$, exhibit the formation of BFS in both spin sectors $\sigma=\;\uparrow, \downarrow$ (Fig.~\ref{fig2}(f)) with anisotropic, sign-alternating energy shifts due to the modified exchange field $\Lambda_{\bm{k};\bm{q}}$ (Fig.~\ref{fig2}(g–h)), similar to the $s$-wave state (Fig.~\ref{fig2}(d–e)). However, in contrast to the $s$-wave state, the finite momentum pairing through $\Delta^{\textrm{FF}}_{\uparrow\downarrow; \bm{q}}$ results in the distorted pattern breaking the reflection symmetry across the line $k_x=-k_y$ or $k_x=k_y$ in the energy spectra.

\begin{figure}[t!]
    \centering
    \includegraphics[width=1\linewidth]{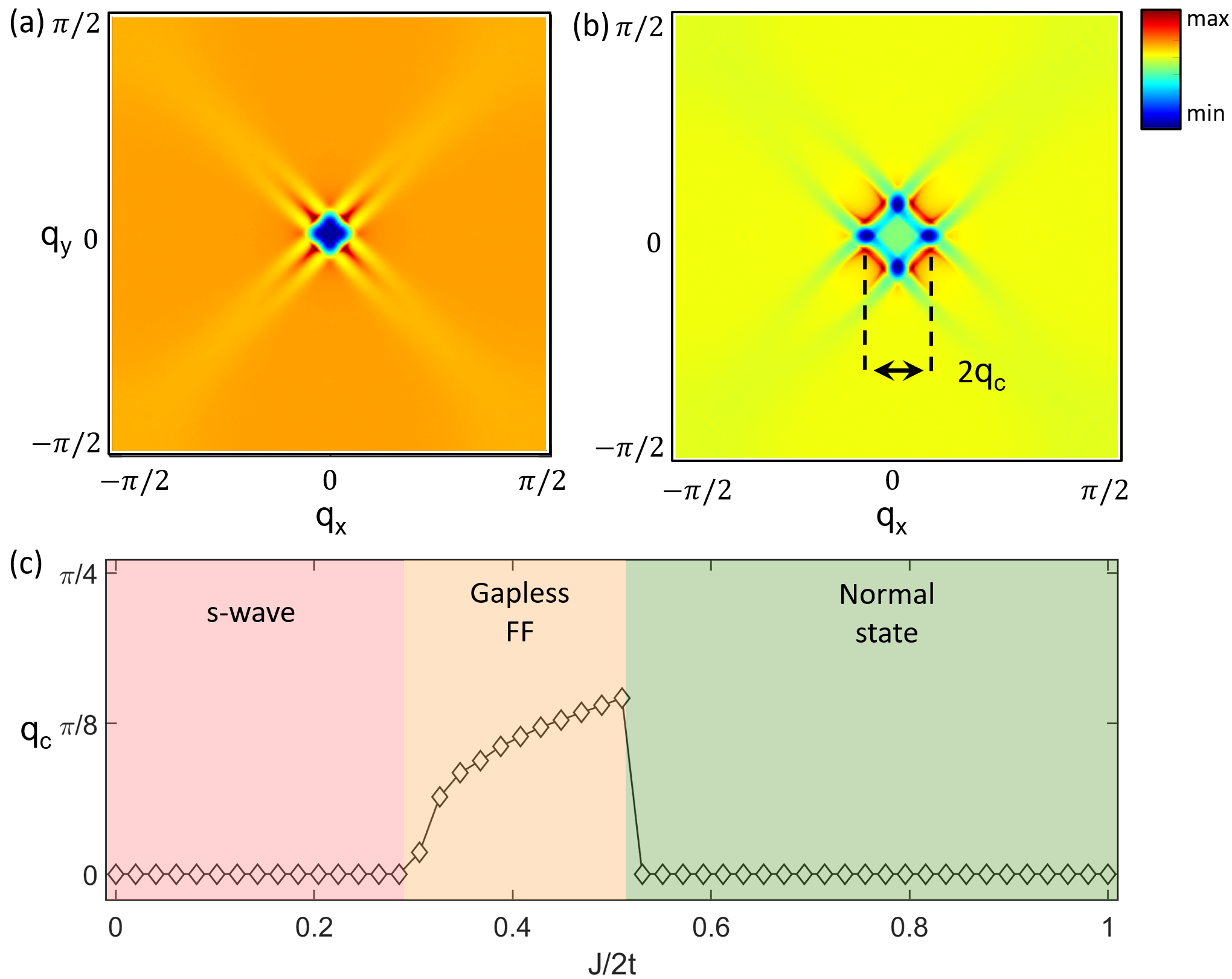}
    \caption{(a–b) The FF state ground state energy ($F_0^{\textrm{FF}}(\bm{q})$) is shown as a function of the pairing momentum vector $\bm{q}=(q_x,q_y)$. (c) Evolution of the optimal pairing momentum $\bm{q}_c$ that minimizes $F_0^{\textrm{FF}}(\bm{q})$. The parameter values $J/2t = 0.2$ and $J/2t = 0.4$ are utilized in (a) and (b), respectively. In all panels, the parameters $V/2t=0.7$, $U/2t=1.5$, and $\mu/2t=-1$ are utilized.}
    \label{fig3}
\end{figure}

An explanation for the dominance of the FF state over the $s$-wave state is the broader overlap between the dispersions of spin-up electrons ($\xi_{\bm{k}+\bm{q}}+J_{\bm{k}+\bm{q}}$) and spin-down holes ($-\xi_{-\bm{k}+\bm{q}}+J_{-\bm{k}+\bm{q}}$), pivotal for finite momentum pairing $c_{\bm{k}+\bm{q}\uparrow}c_{-\bm{k}+\bm{q}\downarrow}$ (the marked region in Fig.~\ref{fig1}(b)). Within this overlapped region, the `energy cost' stemming from the modified dispersion diminishes due to the overlap (i.e., $\Xi_{\bm{k};\bm{q}}=\frac{1}{2}(\xi_{\bm{k}+\bm{q}}+J_{\bm{k}+\bm{q}}+\xi_{\bm{k}-\bm{q}}-J_{\bm{k}-\bm{q}}) \approx 0 $) (see the corresponding equation in Eq.~\eqref{eq:gap_eq}). This extended overlap enhances the contribution to the FF state's pairing amplitude $\Delta^{\textrm{FF}}_{\uparrow\downarrow; \bm{q}}$, resulting in an increased value of $\Delta^{\textrm{FF}}_{\uparrow\downarrow; \bm{q}}$ despite the presence of the BFS. In this scenario, the specific $\bm{q}$ vector characterizing the FF state is determined by optimizing available channels for Cooper pairing. Our analysis reveals that there are four possibilities for the direction of $\bm{q}$: $\pm \hat{x}$ and $\pm \hat{y}$, all of which exhibit degenerate energy levels (Fig.~\ref{fig3}(b)). It is noteworthy that these $\bm{q}$ vectors align with the hopping directions, pointing along one of the lattice vector directions such as $\pm \hat{x}$ and $\pm \hat{y}$. Furthermore, our study shows a rising trend in the magnitude of the $\bm{q}$ vector as the exchange field $J_{\bm{k}}$ increases (Fig.~\ref{fig3}(c)).

The $C_{4z} \mathcal{T}$ symmetry inherent in the model indicates the presence of four-fold degenerate ground state manifolds, characterized by $\bm{q}=\pm(q_{0},0), \pm (0,q_{0})$. In our investigation, we choose one configuration among these four possibilities. However, in principle, one can compose various linear superpositions of the four FF states with different $\bm{q}$, which possibly have lower energy states. Among these compositions, the only possible superconducting order parameter which preserves $C_{4z}\mathcal{T}$ is multiple q-state, which is given as $\Delta \sim \Delta_{(+q_0,0)}+\Delta_{(-q_0,0)}+\Delta_{(0,+q_0)}+\Delta_{(0,-q_0)}$.  While the single-q FF state preserves the effective translational symmetry with the gauge transformation, the multiple-q FF state gives rise to a pair-density wave that breaks the translational symmetry. The investigations on the finer energetic splitting can elucidate the energy differentials between these pair-density waves \cite{sim2024pairdensitywavessupercurrent}.

Upon entering the strong field regime ($J/2t \gg 0.28$), we observe the formation of the $p+ip$ phase (the blue region in Fig.~\ref{fig2}(a)). In this phase, the BdG quasiparticle energy spectra, $E^{\textrm{p+ip}}_{\bm{k}\sigma} = \sqrt{(\xi_{\bm{k}} + s(\sigma) J_{\bm{k}})^{2}+|\Delta_{\sigma\sigma}^{\textrm{p+ip}}g_{\bm{k}}^{\textrm{p+ip}}|^{2}}$, maintain a fully gapped structure (Fig.~\ref{fig2}(i)), regardless of the field strength; except for the case $\mu/2t=0$, where the energy spectra possess nodal points at $(k_x,k_y)=(\pi,0),(0,\pi)$. The preserved gap structure ensures a significant pairing amplitude $\Delta_{\sigma\sigma}^{\textrm{p+ip}}$ (Fig.~\ref{fig2}(i)), as most of the BdG quasiparticle states contribute to $\Delta_{\sigma\sigma}^{\textrm{p+ip}}$ in contrast to the $s$-wave and FF states (see Eq.~\eqref{eq:gap_eq}). Consequently, effective energy minimization is achieved through the gap opening throughout the entire normal-state Fermi surface, establishing the $p+ip$ phase as a stable phase in the strong field regime. The influence of the exchange field primarily results in anisotropic energy spectra without the formation of the BFS (Fig.~\ref{fig2}(j–k)).

We attribute the emergence of the $p+ip$ phase in the strong field regime ($J/2t\gg1$) to its persistent pair correlation function, $(g^\textrm{p+ip}_{\bm{k}})^{*} \left \langle c_{\bm{k}\sigma}c_{-\bm{k}\sigma} \right\rangle$. In the $p+ip$ phase, the absence of BFS leads to the preservation of the pair correlation function across most of the normal-state Fermi surface, excluding at nodal points where the corresponding square harmonics $g^\textrm{p+ip}_{\bm{k}}$ vanishes (Fig.~\ref{fig1}(f)). This stands in contrast with the pair correlation function of the $s$-wave state, $\left \langle c_{\bm{k}\uparrow}c_{-\bm{k}\downarrow} \right\rangle$, which excludes contributions from occupied BdG quasiparticle states, pertaining only four nodal points where the exchange field vanishes (Fig.~\ref{fig1}(d)). With increasing field strength, the presence of occupied BdG quasiparticle states becomes more pronounced. Therefore, in the strong field regime, the $p+ip$ phase emerges as the most energetically favored state due to the amplified exclusion of contributions from these states to the pairing amplitude.

It is important to note that when \(J/2t \gtrsim 0.5\) and \(V/2t \lesssim 0.5\) (indicated by the green area in Fig.~\ref{fig2}(a)), the self-consistent gap equations presented in Eq.~\eqref{eq:gap_eq} do not yield non-zero solutions for pairing amplitudes. This occurs because the energy cost associated with forming Cooper pairs exceeds the energy gain from gapping out the normal metallic band structures for any value of the pairing amplitudes. Consequently, we conclude that a non-superconducting normal metallic state is stabilized in this regime, at least within the framework of our model, which employs single harmonic components for each superconducting state.

\subsection{Stability of gapless FF states}

It is important to evaluate the superfluid density of the FF states since the formation of the BFS potentially makes the value of the superfluid density negative, which indicates the thermodynamic instability of this state \cite{PhysRevB.57.8566, PhysRevB.102.064504}. Employing the following formula \cite{Peotta2015, PhysRevA.74.063626}
\begin{equation}
    [D_{s}]_{i,j} = \frac{\partial^{2} F(\bm{q})}{\partial q_{i} \partial q_{j}}\bigg|_{\bm{q}=\bm{q_{c}}}
\end{equation}
we calculate the superfluid density tensor at zero-temperature, $[D_{s}]_{i,j}$ with $i,j\in\{x,y\}$, for the $s$-wave and FF states. Here, $\bm{q_{c}}$ represents the stationary point of the ground state energy $F(\bm{q})$: for the $s$-wave, $\bm{q_{c}}=\bm{0}$, whereas for the FF state, $\bm{q_{c}}\neq\bm{0}$. Our results demonstrate that the diagonal components of $[D_{s}]_{i,j}$, $[D_{s}]_{x,x}$ and $[D_{s}]_{y,y}$, remain positive-definite even though they tend to decrease with \(J\) throughout the gapless FF phase despite decreasing tendency with \(J\) across the gapless FF phase (the off-diagonal component $[D_{s}]_{x,y}$ remains zero). These findings suggest that the gapless FF state remains thermodynamically stable against the formation of BFS. In contrast, the reduction in $[D_{s}]_{x,x}$ and $[D_{s}]_{y,y}$ with increasing \(J\) signifies a weakening of superconductivity, which includes diminished critical supercurrents and magnetic fields \cite{tinkham2004introduction}. Further details can be found in the SM \cite{SM}.

\subsection{Unconventional superconductivity induced by altermagnetism}

It is worth mentioning that while we primarily focus on the $s$-wave, FF, and $p+ip$ states out of the five potential superconducting states considered in this study, the other two states—the extended $s$-wave and $d$-wave states—may emerge when $V/2t$ is sufficiently large. This behavior aligns with previous studies on the extended Hubbard model for cuprates \cite{PhysRevB.37.9410}. Moreover, our investigation reveals that these states can become prevalent when the exchange field is small ($J/2t \ll 1$) and the chemical potential $\mu$ is tuned to $ \mu/2t = -2 $ or  $\mu/2t = 0 $. For additional details about our analysis, refer to the SM \cite{SM}.

While our focus in this study is on `pure harmonic states,' which correspond to specific types of square harmonics such as \( s \), \( es \), \( d \), \( p + ip \), and \( p - ip \), it is important to note that superconducting states can generally exist as linear combinations of these square harmonics. To assess the impact of these superpositions, we have investigated `mixed ($s$, $es$) states,' which are linear combinations of \( s \)-wave and \(es\)-wave states. Our finding indicates that while the inclusion of the \( es \)-wave component may potentially reduce the area of the FF phase, the FF state still manifests in the phase diagram for sufficiently large chemical potential ($\mu\gtrsim -0.5$), where the \( es \)-wave component is negligibly small compared to the \( s \)-wave component. Additionally, our exploration of `mixed \( p \)-wave states,' which are linear combinations of \( p + ip \) and \( p - ip \), reveals that the mixed $p$-wave states are energetically unfavorable than `pure \( p \)-wave states,' such as \( p + ip \) and \( p - ip \). These findings support the stability of the phase diagram presented in Fig.~\ref{fig2}(a), particularly with regard to the presence of the FF and chiral \( p \)-wave states in this parameter regime. Additional details can be found in the SM \cite{SM}.

The stabilization of the $p+ip$ state over conventional BCS states indicates a promising avenue toward the realization of topological chiral superconductivity. These chiral states possess equal chirality in both spin sectors and a net Chern number of \( C=2 \) for the \( p+ip \) type and \( C=-2 \) for the \( p-ip \) type, exhibiting symmetry-protected chiral edge modes \cite{PhysRevB.108.184505}. Another spin-triplet state of the \( p+ip \) type, known as the `helical \( p \)-wave state,' displays opposite chiralities between the spin-up and spin-down sectors and has a vanishing Chern number \( C=0 \) \cite{PhysRevB.108.184505}. While \( C=0 \) indicates the absence of chiral edge modes, this state instead exhibits protected corner states. It is worth noting that although the degeneracy exists between the chiral states within our model, the presence of weak $C_{4z}\mathcal{T}$-breaking Rashba spin-orbit coupling energetically favors the helical state over the chiral state \cite{PhysRevB.108.184505}.

There have been reports on the discovery of the finite momentum pairing states in the altermagnet. We emphasize the distinctive feature of our observed FF state with BFS, distinguishing it from instances of gapless superconductivity without finite momentum pairing \cite{PhysRevB.109.L201404} and FF states without BFS \cite{PhysRevB.110.L060508, Zhang2024} reported in previous studies. It is also worth mentioning that the observed FF state occurs with $s$-wave pairing, which is different from the $d$-wave FF state reported in Ref.~\cite{PhysRevB.110.L060508}. The difference occurs due to the range of the chemical potential.

\subsection{Experimental signatures}

\begin{figure}
    \centering
    \includegraphics[width=\linewidth]{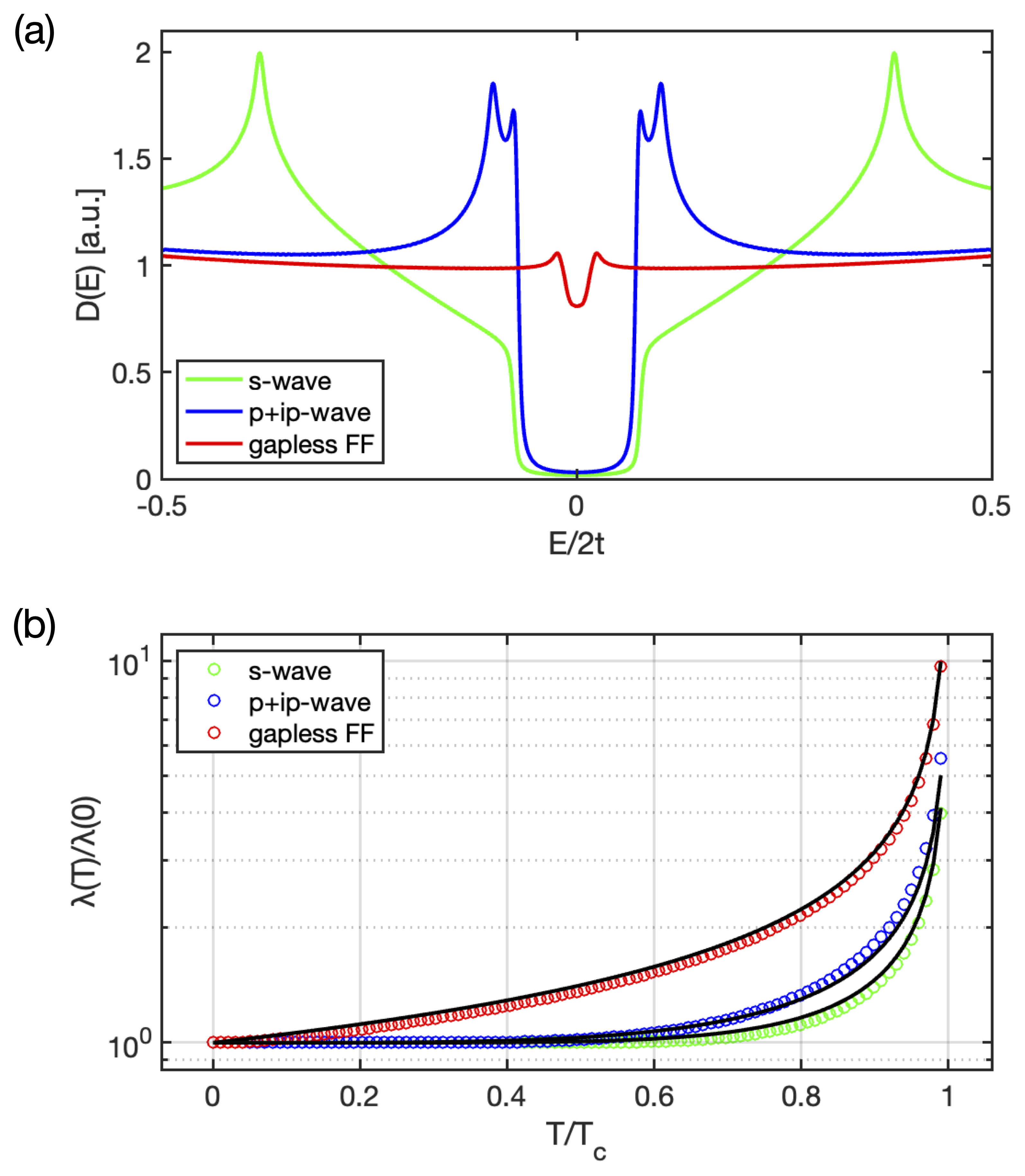}   
    \caption{(a) The density of states, $D(E)$, of the BdG quasiparticle energy at three different states. The green, blue, and red curves represent the $s$-wave, $p+ip$, and gapless FF states, respectively. (b) The normalized London penetration depth, $\lambda(T)/\lambda(0)$, as a function of $T/T_{c}$, where \(T\) is the temperature and \(T_c\) is the critical temperature for each superconducting state. The black lines represent fitting curves described by \(1/\sqrt{1-x^6}\), \(1/\sqrt{1-x^4}\), and \(1/\sqrt{1-x}\), for the \(s\)-wave, \(p+ip\), and gapless FF states, respectively, with \(x = T/T_c\). The self-consistent solution for each state is computed using parameter values \(U/2t = 1.5\), \(V/2t = 1\), and \(\mu/2t = -1\) are applied to across all states. The values \(J/2t = 0.2\), \(J/2t = 0.4\), and \(J/2t = 0.8\) are used for the $s$-wave, FF, and $p+ip$ states, respectively.}
    \label{fig4}
\end{figure}

The density of states (DOS) of the \(s\)-wave, FF, and \(p+ip\) states exhibits distinct characteristics in their gap structures, as illustrated in Fig.~\ref{fig4}(a). The DOS of the \(s\)-wave state shows a fully gapped structure, with coherence peaks shifted away from the edge of the energy gap, unlike typical cases without exchange fields \(J_{\bm{k}}\) (green curve). This shift is attributed to the spin-splitting effects of the exchange fields \(J_{\bm{k}}\), and its magnitude is proportional to the strength of \(J_{\bm{k}}\). In contrast, the DOS of the \(p+ip\) state retains coherent peaks at the edge of the energy gap while exhibiting a splitting into two subpeaks within each coherence peak (blue curve). The gapless FF phase displays a pseudogap structure due to the formation of the BFS (red curve). These distinct features provide clear signatures for discerning these states in experimental DOS measurements, such as scanning tunneling spectroscopy.

We calculate the London penetration depth for the \(s\)-wave, FF, and \(p+ip\) states, following the methodology outlined in Ref.~\cite{tinkham2004introduction}. The London penetration depth \(\lambda(T)\) exhibits distinct characteristics among these three states, as illustrated in Fig.~\ref{fig4}(b). The \(p+ip\) state follows the typical form \(\lambda(T) \simeq \lambda(0)/\sqrt{1-(T/T_c)^4}\) (blue markers), an empirical formula commonly used to describe the London penetration depth in conventional superconductors \cite{tinkham2004introduction}. In contrast, the \(s\)-wave and FF states show significant deviations from this trend (green and red markers, respectively). To fit these data, we introduce \(\lambda(T) \simeq \lambda(0)/\sqrt{1-(T/T_c)^6}\) for the \(s\)-wave state and \(\lambda(T) \simeq \lambda(0)/\sqrt{1-(T/T_c)}\) for the gapless FF state, which fit the data remarkably well. These differences are attributed to the anomalous energy spectra of the \(s\)-wave and gapless FF states, as shown by their DOS in Fig.~\ref{fig4}(b), and can serve as clear indicators for distinguishing these states from the \(p + ip\) state during London penetration depth experiments.

\section{Discussion}

Our investigation into the unconventional superconductivity of two-dimensional altermagnetic metals has uncovered intriguing possibilities beyond conventional BCS states. Through rigorous theoretical analysis, we have elucidated the emergence of topological $p\pm ip$ states, BFSs, and FFLO phase triggered by the interplay between exchange interactions and nearest-neighbor attractive interactions. The occurrence of the BFS plays a crucial role in significantly suppressing conventional BCS superconductivity, thereby enabling the realization of these alternative unconventional superconducting states. We find that in the altermagnetic model, p+ip superconductivity can be stabilized with relatively small nearest neighbor interactions compared with the conventional metallic Fermi surface without spin splitting. This phenomenon offers a promising avenue towards the realization of topological chiral superconductivity.

The observed superconductivity in RuO\textsubscript{2} occurs in the presence of strain that explicitly breaks the underlying $C_{4z}\mathcal{T}$ symmetry. While the four-fold symmetric ground state manifold of the FFLO instability is reminiscent of the original rotational symmetry, the application of the strain would lift the ground state manifold, which would allow the strain engineering of the superconducting pairing symmetry. The study on the effect of the external rotational-symmetry breaking strain would be an interesting topic for future study. \\

\newpage
\section*{Acknowledgement}
M.J.P. thanks Changyoung Kim and Gibaik Sim for the fruitful discussions. This work was supported by the National Research Foundation of Korea (NRF) grant funded by the Korea government (MSIT) (Grants No. RS-2023-00218998). K.-M.K. was supported by the Institute for Basic Science in the Republic of Korea through the project IBS-R024-D1. This work was supported by the BK21 FOUR (Fostering Outstanding Universities for Research) program through the National Research Foundation (NRF) funded by the Ministry of Education of Korea.

\bibliography{reference}

\pagebreak

\end{document}